\documentclass[proceedings, preprint]{rmaa}

\usepackage{paralist}

\usepackage{psfrag,color}

\SetYear{2011}
\SetConfTitle{XIII Latin American Regional IAU Meeting}

\title{GTC long-slit spectroscopy of compact stellar clusters in M81} 

\author{
  Y. D. Mayya,\altaffilmark{1}
  D. Rosa-Gonz\'alez\altaffilmark{1}
  and M. Santiago-Cort\'es\altaffilmark{1}} 

\altaffiltext{1}{Instituto Nacional de Astrof\'\i{}sica,
\'Optica y Electr\'onica, Apartado Postal 51 y 216, Puebla, M\'exico
 (ydm, danrosa, scortes@inaoep.mx).}

\shortauthor{Mayya, Rosa-Gonz\'alez, \& Santiago-Cort\'es}
\shorttitle{GTC spectra of M81 Compact Star Clusters}

\listofauthors{Y. D. Mayya, D. Rosa-Gonz\'alez, \& M. Santiago-Cort\'es}
\indexauthor{Mayya, Y. D.}
\indexauthor{Rosa-Gonz\'alez, D.}
\indexauthor{Santiago-Cort\'es, M.}

\abstract{
We here present the ages of four compact stellar clusters (CSCs) in the nearby 
spiral galaxy M81, using long-slit optical spectra  
obtained with the 10.4-m Gran Telescopio Canarias (GTC). 
All the four CSCs, including the brightest in this galaxy, are found to have ages
between 5--6~Myr, with one of them showing Wolf-Rayet spectral features.
The photometric masses of these clusters, calculated using their 
spectroscopically-derived ages, lie between 3000--18000~M$\odot$. 
The observed clusters are among the brightest objects, and hence the most 
massive, in the entire disk of M81. 
This implies the absence of massive ($10^5$~M$\odot$) compact stellar clusters in M81.
}

\resumen{
En este trabajo presentamos edades de 4 c\'umulos estelares compactos de la gal\'axia
cercana M81. Las edades fueron determinadas usando los espectros de rendija larga
obtenidos con el Gran Telescopio Canarias (GTC). Obtuvimos edades entre 5--6~Myr
para los 4 c\'umulos, incluyendo el c\'umulo m\'as brillante de M81. Uno de los
c\'umulos muestra caracteristicas espectrales de estrellas Wolf-Rayet.
Las edades obtenidas sugieren masas entre 3000--18000~M$\odot$. 
Los objectos observados fueron los m\'as
brillantes, y por lo tanto los m\'as masivos, de toda la galaxia.
Esto implica la ausencia de c\'umulos compactos masivos ($10^5$~M$\odot$) en M81.
}

\addkeyword{galaxies: individual (M81)}
\addkeyword{galaxies: star clusters}
\addkeyword{telescopes: Gran Telescopio Canarias (GTC)}

\begin{document}
\maketitle

\section{Introduction and observations}
\label{sec:intro}

The similarity in the ranges of sizes and masses of the compact star clusters (CSC) 
and the globular clusters (GCs) has given rise to the notion of an
evolutionary connection between them \citep{deG07}. The vast difference in the ages of the
two populations, and the absence of any prototypical cluster of age intermediate
between these two extreme cases, has prevented progress in exploring further this idea.
Crucial for understanding the evolutionary connection between these two populations
is the identification of compact clusters of intermediate ages ($\sim10^8$~yr).
Spectroscopic observations are vital to ascertain ages in these age ranges. 

The nearby galaxy M81 offers a great opportunity to address the evolutionary 
connection, as this galaxy contains both the CSCs and GCs in large numbers (Chandar
et al. 2001; Santiago-Cortes et al. 2010). Santiago-Cortes et al. (2010),
using 29 HST/ACS fields cataloged 263 CSCs brighter than B=22 mag, and 
172 GCs. The photometric masses of the CSCs were found to be less than $2\times10^4$~M$\odot$, 
assuming that the brightest clusters are younger than 10~Myr. According to a recent
study by \citet{Bastian2008}, the highest cluster mass in a galaxy depends on its 
star formation rate (SFR). The expected highest mass of the cluster for the observed
SFR of 1 M$\odot$\,yr$^{-1}$ in M81 is around a factor of 5 greater than that inferred.
On the other hand, if the brightest clusters are older than 100~Myr, observationally
inferred maximum cluster mass would follow the relation suggested by \citet{Bastian2008}.
We hence carried out spectroscopic observations of the brightest clusters with the
goal of determining their ages, and therein, determine their masses.

\section{Observations and data reductions}

Spectroscopic observations were carried out using the long-slit 
of the spectrograph of the OSIRIS instrument at the 10.4-m GTC
\footnote{{\it Gran Telescopio Canarias} is a Spanish initiative with the participation
of Mexico and the US University of Florida, and is installed at the Roque de los Muchachos 
in the island of La Palma. This work is based on the proposal GTC11$-$10AMEX$\_$0001.} 
in the service mode on 2010 April 4 and 5.
Six slit positions were used to obtain spectra of 13 clusters brighter
than $B=21$~mag (11 CSCs, including the brightest CSC, and 2 GCs), 
and a few fainter CSCs.
Spectra cover a range of 3630 to 7500~\AA, at a spectral resolution of $\sim7$~\AA.
A slit-width of 1.0~arcsec was used. The estimated seeing during these observations 
is $\sim1$~arcsec.

The data reduction was carried out in the standard manner using the tasks
available in the IRAF software package. 
The spectra were extracted so as to include all the
observed H$\alpha$ emission associated to the cluster in the spatial direction.
Extracted spectra of four of the clusters are shown in Figure~1.
\begin{figure}[t]
 \centering
  \vskip -3mm
  \includegraphics[width=1.1\columnwidth]{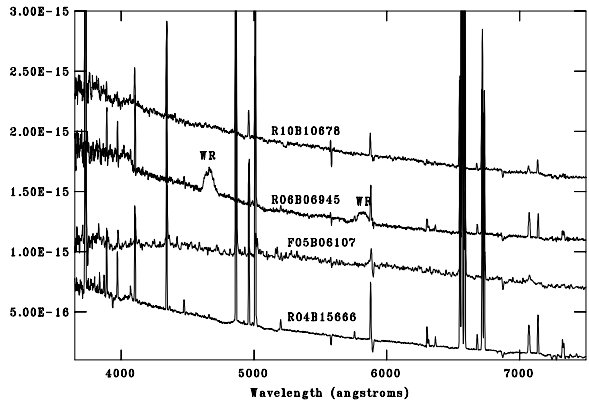}
  \vskip -3mm
\caption{Extracted spectra of 4 compact stellar clusters. The fluxes are
in units of erg\,cm$^{-2}$\,s$^{-1}$\AA$^{-1}$. The fluxes of the top 
3 spectra are displaced upwards for the sake of clarity. }
  \label{fig:fig1}
\end{figure}
\begin{figure}[t]
  \vskip -6mm
  \includegraphics[width=\columnwidth]{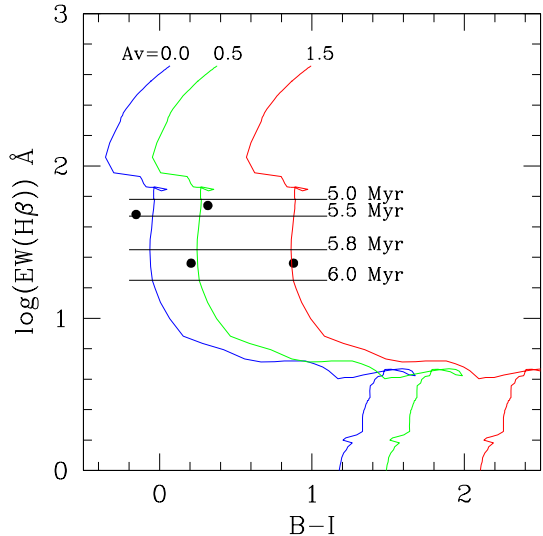}
  \vskip -3mm
\caption{Age and reddening of M81 clusters. Evolutionary track for an 
instantaneous burst model using STARBURST99 (Leitherer et al. 1999)
is shown, rededdened by $Av=0, 0.5$ and $1.5$ magnitudes. The horizontal lines
denote the EW at 4 epochs between 5 and 6~Myr.}
  \label{fig:fig3}
\end{figure}

\section{Results}

All the targeted CSCs are found to be associated with nebular emission. In this
work, we show the results for the brightest CSC, R04B15666, and three other CSCs,
all brighter than B=20~mag. The H$\beta$ emission equivalent widths (EW) measured on 
the extracted spectra are plotted against their $F435W - F814W$ ($B-I$ for short) color 
in Figure~2. It can be seen that all the clusters
are consistent with ages between 5 and 6~Myr, with extinction $Av<1.5$~mag.

Considering that the 1~arcsec slit-width (18~pc) encloses a region larger than the
cluster size $<8$~pc (see the contribution of Santiago-Cort{\'e}s et al. in this volume), 
the presence of emission lines at the position of the cluster 
does not necessarily imply that the cluster stars are responsible for the ionization. 
Massive stars in the field and associations around the cluster could as well
be responsible for the ionization. The
fool-proof way of establishing that the clusters are young objects is by 
identifying age-sensitive spectral features originating in the atmospheres of
cluster stars. All the four spectra show rising continuum 
in the blue, without any signs of absorption features typical of intermediate-age or old
stellar populations, as can be seen in Figure~1. 
One of the studied clusters, R06B06945, clearly shows broad
emission lines, which are denoted by the letters WR in the figure. 
We identify these as CIII~$\lambda4650$ and CIV~$\lambda5806$ features,
characteristic of Wolf-Rayet stars of type WC. The age of
this cluster as inferred from EW is consistent with the presence of WR features.
Thus, our spectra confirm the young age of the brightest CSCs. We obtain masses
between 3000--18000~M$\odot$ for the 4 CSCs, all below the upper mass limit of
$2\times10^4$~M$\odot$ derived by Santiago-Cort\'es et al. (2010).

\section{CONCLUSIONS}

We have carried out spectroscopic observations of 11 CSCs and 2 GCs using the
10.4-m GTC, and present here the results for 4 CSCs, including that for the brightest CSC. 
The H$\beta$ emission equivalent width, optical colors from HST, and the absence of stellar
absorption lines in the blue, together suggest ages less than 6~Myr for the clusters. In one of these 
CSCs, we detected WR features characteristic of WC stars. The determined young age of these
bright objects, which lie in the upper part of the cluster color-magnitude diagram, ensures 
that there are no CSCs in M81 with masses exceeding $2\times10^4$~M$\odot$. Thus, its seems
that though compact clusters are formed in large numbers in star-forming galaxies such as M81,
conditions in these giant galaxies don't favor the formation of massive clusters.

\end{document}